\documentclass[aps,pra,twocolumn,superscriptaddress]{revtex4}
\usepackage[intlimits]{amsmath}
\usepackage{amsfonts}
\usepackage{graphics}
\usepackage{subfigure}
\usepackage[usenames]{color}
\usepackage{epstopdf,epsfig}
\usepackage{color}
\usepackage{indentfirst}
\usepackage{graphicx}

\usepackage{bbold} 

\usepackage[usenames]{color}
  \expandafter\let\csname equation*\endcsname\relax
  \expandafter\let\csname endequation*\endcsname\relax
 \usepackage{mathdots}
 \usepackage{dsfont}

\newcommand{\bra}[1]{\left< #1 \right|}

\newcommand{\ket}[1]{\left| #1 \right>}

\newcommand{\nin}{\in \hspace{-1.8mm} / \,}
\def\N {{\mathds N}}
\def\id {{\mathds 1}}

\begin{document}

\title{Transfer of arbitrary two qubit states via a spin chain}

\author{S. Lorenzo}
\affiliation{Dipartimento di Fisica e Chimica, Universit\`{a} degli Studi di Palermo, Via Archirafi 36, I-90123 Palermo, Italy}

\author{T. J. G. Apollaro}
\affiliation{NEST, Scuola Normale Superiore \& Istituto di Nanoscienze-CNR, I-56126 Pisa, Italy}
\affiliation{Dipartimento di Fisica e Chimica, Universit\`{a} degli Studi di Palermo, Via Archirafi 36, I-90123 Palermo, Italy}
\affiliation{Centre for Theoretical Atomic, Molecular, and Optical Physics, School of Mathematics and Physics, Queen's University Belfast,BT7,1NN, United Kingdom}
\affiliation{International Institute of Physics, Universidade Federal
do Rio Grande do Norte, 59078-400 Natal-RN, Brazil}

\author{Simone Paganelli}
\affiliation{International Institute of Physics, Universidade Federal
do Rio Grande do Norte, 59078-400 Natal-RN, Brazil}
\affiliation{Dipartimento di Scienze Fisiche e Chimiche, Universit\`{a} dell'Aquila, via Vetoio, I-67010 Coppito-L'Aquila, Italy}

\author{G. M. Palma}
\affiliation{Dipartimento di Fisica e Chimica, Universit\`{a} degli Studi di Palermo, Via Archirafi 36, I-90123 Palermo, Italy}
\affiliation{NEST, Scuola Normale Superiore \& Istituto di Nanoscienze-CNR, I-56126 Pisa, Italy}

\author{F. Plastina}
\affiliation{Dip. Fisica, Universit\`{a} della Calabria, 87036 Arcavacata di Rende (CS), Italy}\address{INFN - Gruppo Collegato di Cosenza}


\begin{abstract}
\noindent
We investigate the fidelity of the quantum state transfer (QST) of two qubits by means of an arbitrary spin-$\frac{1}{2}$ network, on a lattice of any dimensionality. Under the assumptions that the network Hamiltonian preserves the magnetization and that a fully polarized initial state is taken for the lattice, we obtain a general formula for the average fidelity of the two qubits QST, linking it to the one- and two-particle transfer amplitudes of the spin-excitations among the sites of the lattice. We then apply this formalism to a 1D spin chain with $XX$-Heisenberg type nearest-neighbour interactions adopting a protocol that is a generalization of the single qubit one  proposed in Ref.~[Phys. Rev. A {\bf{87}}, 062309 (2013)]. We find that a high-quality two qubit QST can be achieved provided one can control the local fields at sites near the sender and receiver. Under such conditions, we obtain an almost perfect transfer in a time that scales either linearly or, depending on the spin number, quadratically with the length of the chain.
\end{abstract}
%
%
%
%
%

\maketitle
\section{Introduction}\label{S.Intro}
The capability of faithfully transferring information from one
location to another is one of the main driving factors of the
modern technological progress. As far as classical information is
concerned, there is no limit, at least in principle, to reproduce
an exact copy of the original message; therefore the information
transfer has to face somehow minor problems than those faced in
the quantum realm. There, the no-cloning
theorem~\cite{WoottersZ82} explicitly prohibits  to make an exact
copy of the quantum state on which the quantum information has
been coded in. This has stimulated, over the last past decades, a
large body of works on how to efficiently achieve Quantum State
Transfer (QST).

For short-haul transfers of the quantum state of a single qubit
(1-QST), the use of spin-$\frac{1}{2}$ chains, initially proposed
in Ref.~\cite{Bose03}, has been largely investigated (see
Refs.~\cite{channels1,channels2} and references therein, and Ref.
\cite{boson} for an implementation with a cavity array). Protocols
based on time-dependent couplings~\cite{timecoupl1,timecoupl2},
fully engineered interactions~\cite{enginereed1,enginereed2},
ballistic
transfer~\cite{ballistic1,ballistic2,ballistic3,ballistic4,ballistic5,ballistic6},
Rabi-like
oscillations~\cite{Rabi-like1,Rabi-like2,Rabi-like3,Rabi-like4,
Rabi-like4a,Rabi-like5,Rabi-like6,Rabi-like7,Rabi-like8,
Rabi-like9,dePasquale05,dePasquale04}, just to name a few, have been shown to achieve high
fidelity 1-QST, in addition to some additional tasks like routing
of the quantum information to an on-demand location on a spin
graph~\cite{routing1,routing2,routing3}.

Recently, the same effort is being devoted to the case of
multiqubit QST ($n{-}$QST), in which the state aimed at being
transferred is made of $n>1$ qubits. In many cases, the adopted
strategies consist of extensions of 1-QST protocols and, as a
consequence, the drawbacks and inconveniences they already
presented for the 1-QST are, to some extent, even more amplified
when it comes to the $n$-QST case.  For example, the multi-rail
scheme~\cite{multirail1,multirail2} requires the use of several
quantum spin-$\frac{1}{2}$ chains and a complex encoding and
decoding scheme of the quantum states; employing linear chains
made of spins of higher dimensionality reduces the number of
chains to one, but still requires a repeated measurement process
with consecutive single site operations~\cite{Bayat14}; the
fully-engineered chain (eventually combined with the ballistic or
Rabi-like mechanism), as well as the uniformly coupled chain with
specific conditions on its length, needs  conditional quantum
gates to be performed on the recipients of the quantum
state~\cite{nPST1,nPST2, nPST3}. Therefore, simpler many qubits
QST schemes would be quite appealing.

In the present paper, we adopt a minimal engineering and
intervention point of view, looking for a 2-QST protocol that does
not need demanding operations to be performed, neither in the form
of external end-operations on the spins nor to engineer the spin
couplings. Experimentally-friendly $2$-QST schemes are interesting
in view of the fact that both the full modulation of the couplings
may be unattainable (depending on the physical system meant to
perform the QST) and that quantum operations, such as measurements
and gates, are prone to errors which, in a realistic set-up, may
fatally degrade the efficiency of the protocol. In addition, as
the exchange of quantum information is meant to occur, for
instance, between quantum processors, it is quite natural that QST
of more than a single qubit has to be faced in order to fully
exploit the potentialities of quantum computation. Needless to
say, also other fields relying on quantum information processing,
such as cryptography and dense coding would widely benefit from
efficient $n$-QST protocols~\cite{BooksQIP}.

The paper is organized as follows: in Sec.~\ref{S.Fidelitygen}, we
obtain a general expression for the average fidelity of the
quantum state transfer of two qubits coupled to an arbitrary total
angular momentum conserving graph of spin-$\frac{1}{2}$
initialized in the fully polarized state; in Sec.~\ref{S.MandP} a
specific one-dimensional instance of such a graph is presented and
it is shown that, by means of strong local magnetic fields on the
so-called {\textit{barrier qubits}}~\cite{Rabi-like6},
high-quality 2-QST can be achieved. Finally, in Sec.~\ref{S.Concl}
some concluding remarks are reported together with a discussion on
possible extensions of our idea.

\section{Fidelity for a class of spin-$\frac{1}{2}$ Hamiltonians}\label{S.Fidelitygen}

In this Section we derive a general expression for the average
fidelity of a 2-qubit quantum state transfer from a pair of
senders to a pair of receivers, residing respectively on sites
$\mathcal{S}=\{s_1,s_2\}$ and $\mathcal{R}=\{r_1,r_2\}$ of a
lattice $\mathcal{K}$ of arbitrary dimensionality. The only
constraints we assume to be satisfied by the spin dynamics on
$\mathcal{K}$ are 1.) the conservation of the total magnetization
$M^{\alpha}=\sum_{n \in {\mathcal{K}}} S_n^{\alpha}$ along some
axes $\alpha$ (which we assume hereafter to be the quantization
axes $z$) and 2.) the initialization of all the spins of
$\mathcal{K}$ but $\mathcal{S}$ into a fully polarized state along
$z$.

The most general Hamiltonian, allowing up to two-body interactions, for spin-$\frac{1}{2}$ particles is given by
\begin{equation}\label{E.Hamgen}
H=\sum_{i j \in \mathcal{K}} \sum_{\alpha \beta} J_{\alpha \beta}^{ij} S_{i}^{\alpha} S_{j}^{\beta}~,
\end{equation}
where $\alpha,\beta=\{0,x,y,z\}$ with $S^0 = \frac{\id}{2}$. Because of
the conservation rule implied by $\left[M^z,H \right]=0$, Eq.
(\ref{E.Hamgen}) can be decomposed into a direct sum over all
subspaces with fixed $z$-component of the angular momentum,
$H=\bigoplus_{S_z} H_{S_z}$,
$S_z={-}\frac{N}{2},-\frac{N}{2}{+}1,..., \frac{N}{2}$. Without
loss of generality we re-scale the labelling of the angular
momentum sectors by the number $n$ of spins flipped in each
sector, that is $S_z=-\frac{N}{2}+n$  with $n=0, 1,2,...,N$. The
Hilbert space dimension of the $n$-th sector is clearly $2^ {N
\choose n}$.

Our goal is to transfer the quantum state of two qubits located at
sites $\mathcal{S}$ and given by
$\ket{\psi(0)}_{\mathcal{S}}{=}\alpha \ket{00}{+}\beta
\ket{01}{+}\gamma \ket{10}{+}\delta \ket{11}$ to the receivers
spin, located at sites $\mathcal{R}$. The rest of the lattice,
embodied by the quantum channel $\Gamma$ and the receivers
$\mathcal{R}$, is initialized in the state $\ket{\Gamma
\mathcal{R}}=\otimes_{j\nin \mathcal{S}}\ket{0}_j$.
The evolution of the overall state $\ket{\Psi(t)}$ is given by
\begin{eqnarray}\label{E.Evogen}
\ket{\Psi(t)}&=&e^{- i H t}\ket{\psi(0)}_{\mathcal{S}}\ket{\Gamma \mathcal{R}}=  \nonumber \\
&=&e^{- i H_0 t}\alpha\ket{00}_{\mathcal{S}} \ket{\Gamma \mathcal{R}}+  \nonumber \\ 
&&+  e^{- i H_1 t}\left(\beta\ket{01}_{\mathcal{S}}{+}\gamma\ket{10}_{\mathcal{S}}\right)\ket{\Gamma \mathcal{R}} + \nonumber  \\
&&+ e^{- i H_2 t}\delta\ket{11}_{\mathcal{S}}\ket{\Gamma \mathcal{R}},
\end{eqnarray}
where the Hamiltonian has been restricted to the
subspaces $n=0,1,2$, respectively, by taking into account the
invariant sector of the Hilbert space to which each component of
the state vector pertains.

By tracing out all of the spins but the receivers, one obtains the
state of the latter, $\rho_{\mathcal{R}}(t)=Tr_{\mathcal{K} -
\mathcal{R}}\left(\ket{\Psi(t)}\!\!\bra{\Psi(t)} \right)$. The
fidelity between the state transferred to the receivers and the
state encoded initially on the senders is given by~\cite{Josza94}
\begin{equation}\label{E.Fidelity}
F\left(\ket{\psi(0)}\!\!\bra{\psi(0)}_{\mathcal{S}},\rho_{\mathcal{R}}(t)\right)={}_{\mathcal{S}}\!\bra{\psi(0)}\rho_{\mathcal{R}}(t)\ket{\psi(0)}_{\mathcal{S}}.
\end{equation}

The quality of a QST protocol, however, cannot be simply evaluated
by considering the fidelity of the  transfer of a single, specific
input state; in fact, a more appropriate figure of merit is given
by the average QST-fidelity $\bar{F}(t)$ obtained by averaging
over all possible input states.

After a lengthy but straightforward calculation, full details are reported in \cite{manyqubits}
we obtain the
average fidelity $\bar{F}(t)$ for the 2-QST with the constraints
of a lattice $\mathcal {K}$ described by a total $z$-magnetization
conserving Hamiltonian and provided the fully polarized initial
state is taken for $\Gamma$ and $\mathcal{R}$ :
\begin{widetext}
\begin{eqnarray}\label{E.Fidelitygen}\hspace{-1.5cm}
\bar{F}(t)&=&\frac{1}{4}{+}\frac{5}{54}Re\left[f_{s_1}^{r_1}{+}f_{s_2}^{r_2}+\frac{7}{5}f_{s_2}^{r_2} (f_{s_1}^{r_1})^* \right]{+}\frac{1}{54}\left(|f_{s_2}^{r_1}|^2{+}|f_{s_1}^{r_2}|^2 \right)
{+}\frac{5}{108}\left( |f_{s_2}^{r_2}|^2{+}|f_{s_1}^{r_1}|^2\right){+}\frac{7}{54}Re\left[g_{s_1 s_2}^{r_1 r_2}\right]{+}
\frac{5}{108}|g_{s_1 s_2}^{r_1 r_2}|^2 + \nonumber \\
&&{-}\frac{1}{54}\left(1{-}\sum_{n{<}m{=}1}^{n,m {\nin}\mathcal{R}} |g_{s_1 s_2}^{nm}|^2 \right)
{+}\frac{5}{54}Re\left[\left(f_{s_1}^{r_1}{+}f_{s_2}^{r_2}\right)(g_{s_1 s_2}^{r_1r_2})^*\right]{-}\frac{1}{27}\sum_{n{=}1}^{n{\nin}\mathcal{R}}Re\left[(f_{s_2}^{n})^*g_{s_1s_2}^{n r_1}{+}(f_{s_1}^{n})^*g_{s_1 s_2}^{nr_2}\right],
\end{eqnarray}
\end{widetext}
where $f_n^m{=}\bra{m}e^{-i t H_1}\ket{n}$ and
$g_{nm}^{rs}{=}\bra{rs}e^{-i t H_2}\ket{nm}$ are the single- and
two-particle transfer amplitudes from sites $n\rightarrow m$ and
$\{nm\}\rightarrow \{rs\}$, respectively.

Eq.~(\ref{E.Fidelitygen}) plays the same role for the 2-QST of the
celebrated average fidelity expression given in Ref.~\cite{Bose03}
for the 1-QST.

Notwithstanding the lengthy expression for the average fidelity,
in the presence of further symmetries and specific Hamiltonians
intended to implement the 2-QST protocol,
Eq.~(\ref{E.Fidelitygen}) can be considerably simplified. In the
next Section we give an instance of such a procedure and, at the
same time, we propose a model that accomplishes a high-quality
2-QST.

\section{The model and the protocol}\label{S.MandP}

The results for the 2-QST scheme we propose in this Section have
to be compared with the average fidelity (hereafter called
fidelity) we would obtain by means of local operations and
classical communication (LOCC) or by means of universal quantum
cloning machines (UQCM). The use of these channels yields what is
conventionally dubbed as classical fidelity and amounts to,
respectively, $F_{\text{LOCC}}{=}\frac{2}{d+1}$~\cite{clasF} and
$F_{\text{UQCM}}{=}\frac{d+2}{2\left(d+1\right)}$~\cite{UQCM}, where $d$ is
the Hilbert-space dimension of the state aimed to be transferred.
For the case of 2 qubits we have $d{=}4$ and, therefore, our
protocol outperforms the classical ones if we obtain a fidelity
higher than $\frac{3}{5}$ (or $\frac{2}{5}$ if optimal cloning is
not available for the system at hand).

The lattice $\mathcal{K}$ we will consider is a 1D
spin-$\frac{1}{2}$ open chain and the Hamiltonian is taken of the
$XX$-Heisenberg type with nearest-neighbor interactions only, and
a magnetic field along the $z$-axis on the 3rd and $(N-2)$th spin,
playing the role of the `barrier' qubits, separating the ${\cal
S}$ and ${\cal R}$ pairs from the rest of the channel:
\begin{eqnarray}\label{E.HamBus}
H=-\sum_{l=1}^{N-1} J_l(\sigma^x_l \sigma^x_{l+1}+\sigma^y_l \sigma^y_{l+1})+h\left(\sigma^z_3+\sigma^z_{N-2}\right),
\end{eqnarray}
where $\sigma^{\alpha}=2 S^{\alpha}$ ($\alpha=x,y,z$) are the usual Pauli matrices.

\begin{figure}[h!]\centering
\includegraphics[width=1.0\columnwidth]{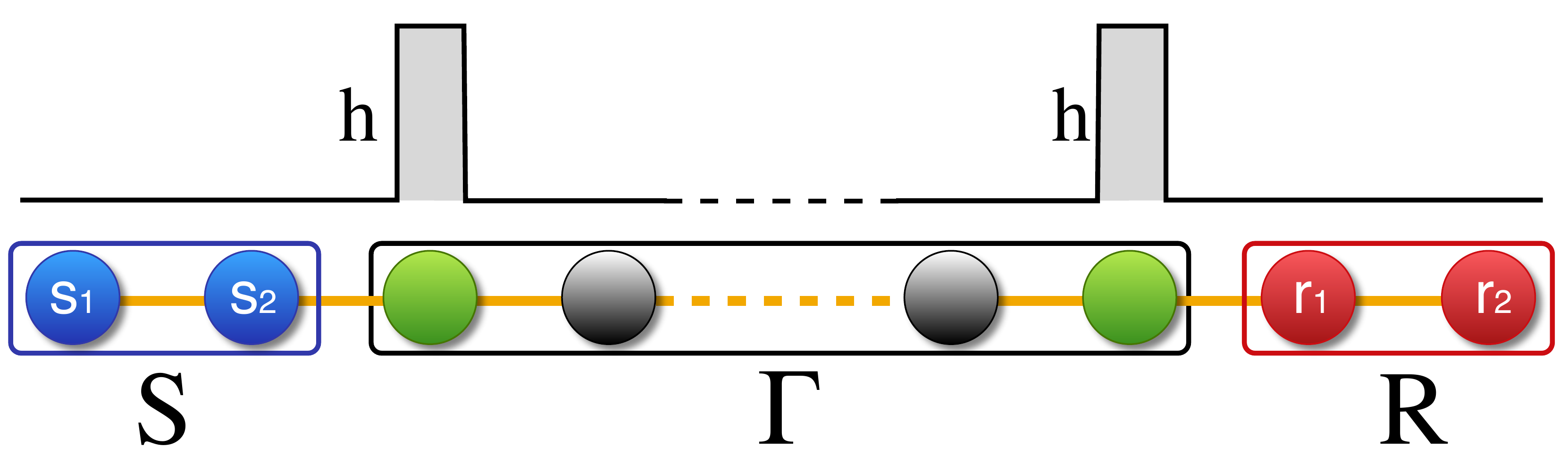}
\caption{(Color online) The spin graph $\mathcal{K}$ by means of
which we aim at achieving high-quality 2-QST via a Rabi-like
mechanism between the two ends of the spin chain. The quantum
state to be sent is encoded on the spins $\mathcal{S}=\{1,2\}$
(blue), whereas the receivers are located at
$\mathcal{R}=\{N-1,N\}$ (red).  Barrier qubits, residing at sites
$n=\{3,N-2\}$, on which a strong magnetic field $h$ is applied are
colored in green. }
  \label{F.1qubitQSTMB}
\end{figure}

We aim to achieve the transfer of an arbitrary  2-qubit state
residing on the sender spins $\mathcal{S}$, located at sites 1 and
2, $\ket{\psi(0)}_{12}{=}\alpha \ket{00}{+}\beta \ket{01}{+}\gamma\ket{10}{+}\delta \ket{11}$, to the receiver spins $\mathcal{R}$,
residing at the other end of the chain, $\mathcal{R}=N-1,N$, as
depicted in Fig.~\ref{F.1qubitQSTMB}.

Eq.~(\ref{E.HamBus}) can be mapped to a tight-binding spinless
fermion model via the Jordan-Wigner
transformation~\cite{LiebSM1961}
\begin{equation}
\label{E.Hfermrealspace}
H=-\sum_{i=1}^{N-1}c^{\dagger}_i c_{i+1} {+}c_i c^{\dagger}_{i+1}{+}h\left(c^{\dagger}_{3} c_{3}{+}c^{\dagger}_{N{-}2} c_{N{-}2}\right),
\end{equation}
where we have taken as our energy and inverse time unit the
exchange energy $J$, that we consider to be site independent.

Because of the quadratic nature of the Hamiltonian, the single
particle spectrum is sufficient to describe the full
dynamics. Denoting by $\varepsilon_k$ and $\ket{\varepsilon_k}$
the $k$-th energy eigenvalue and its corresponding eigenvector,
the full Hamiltonian operator acting on a $2^N$ dimensional
Hilbert space, is easily decomposed into a direct sum over all
particle number-conserving invariant subspaces
$H=\bigoplus_{n=1}^N H_n$, where
\begin{widetext}
\begin{equation}\label{E.decom}
H_n{=}\!\!\!\!\!\!\!\!\sum_{k_1{<}k_2{<}...{<}k_n{=}1}^N
\!\!\!\!\!\!\!\!\!\!\!\!
\left(\varepsilon_{k_1}{+}\varepsilon_{k_2}{+}...{+}\varepsilon_{k_n}\right)
c^{\dagger}_{k_1}c^{\dagger}_{k_2}...c^{\dagger}_{k_n}\ket{\{0\}}\bra{\{0\}}c_{k_n}...c_{k_2}c_{k_1},
\end{equation}
\end{widetext}
with $\ket{\{0\}}$ being the fermion vacuum. Each $H_n$, therefore
can be constructed quite simply once the single-particle spectrum
is known. Notice that the specific ordering of the $k_i$\rq{}s in
the sum of Eq.~(\ref{E.decom}) is taken in such a way that
unwanted phase factors do not arise when mapping back into spin
operators via the inverse Jordan-Wigner transformation.

Therefore, in order to evaluate $\ket{\Psi(t)}$ as given by
Eq.~(\ref{E.Evogen}), we need the spectral resolution of $H_1$,
given by the eigenvalues and eigenvectors of the following $N\times N$
tri-diagonal matrix:
$T_{ij}=\delta_{i,i+1}{+}\delta_{i,i-1}{+}h_{3,3}{+}h_{N-2,N-2}$, which
is easily diagonalizable, at least numerically. Notice that a
uniform magnetic field along the $z$-direction has no influence on
what follows as it corresponds to adding a term proportional to
the Identity in Eq.~\ref{E.Hfermrealspace}. Hence the eigenvectors
do not change, whereas the uniform shift experienced by all of the
eigenvalues is cancelled out in the time evolution of the
fidelity, which, as we will show below, only depends on energy
differences.

Key to our aim is the presence of eigenstates that are at the same
time strongly localized on both the sender and the receiver spins.
That is, by expanding the Hamiltonian eigenvectors in the position
basis $\ket{\varepsilon_k}{=}\sum_{n{=}1}^N a_{kn}\ket{n}$ (where
$\ket{n}\equiv \ket{0_1....0_{n-1} 1_n 0_{n+1}...0_N}$ describes a
state with a single spin flipped at position $n$), a prerequisite
for  Rabi-like oscillations based 2-QST protocol to correctly work
is that there exists a certain (small) number of eigenstates
$\ket{\varepsilon_k}$ for which $a_{kn}$ is non-negligible only
for $n=1,2,N{-}1,N$.
We find this requirement of {\textit{edge quadri-localization}} to
be fulfilled for spin chain of lengths  $N{\neq}3n{-}1$, where
$n\geq3\in \N$, when strong magnetic fields are applied on the
barrier qubits. The condition $n\geq3$ is due to the fact that the
minimum length of a spin chain allowing for 2 senders, 2
receivers, and  2 barrier qubits is $N\geq6$.

\begin{figure}[h!]
\includegraphics[width=\columnwidth]{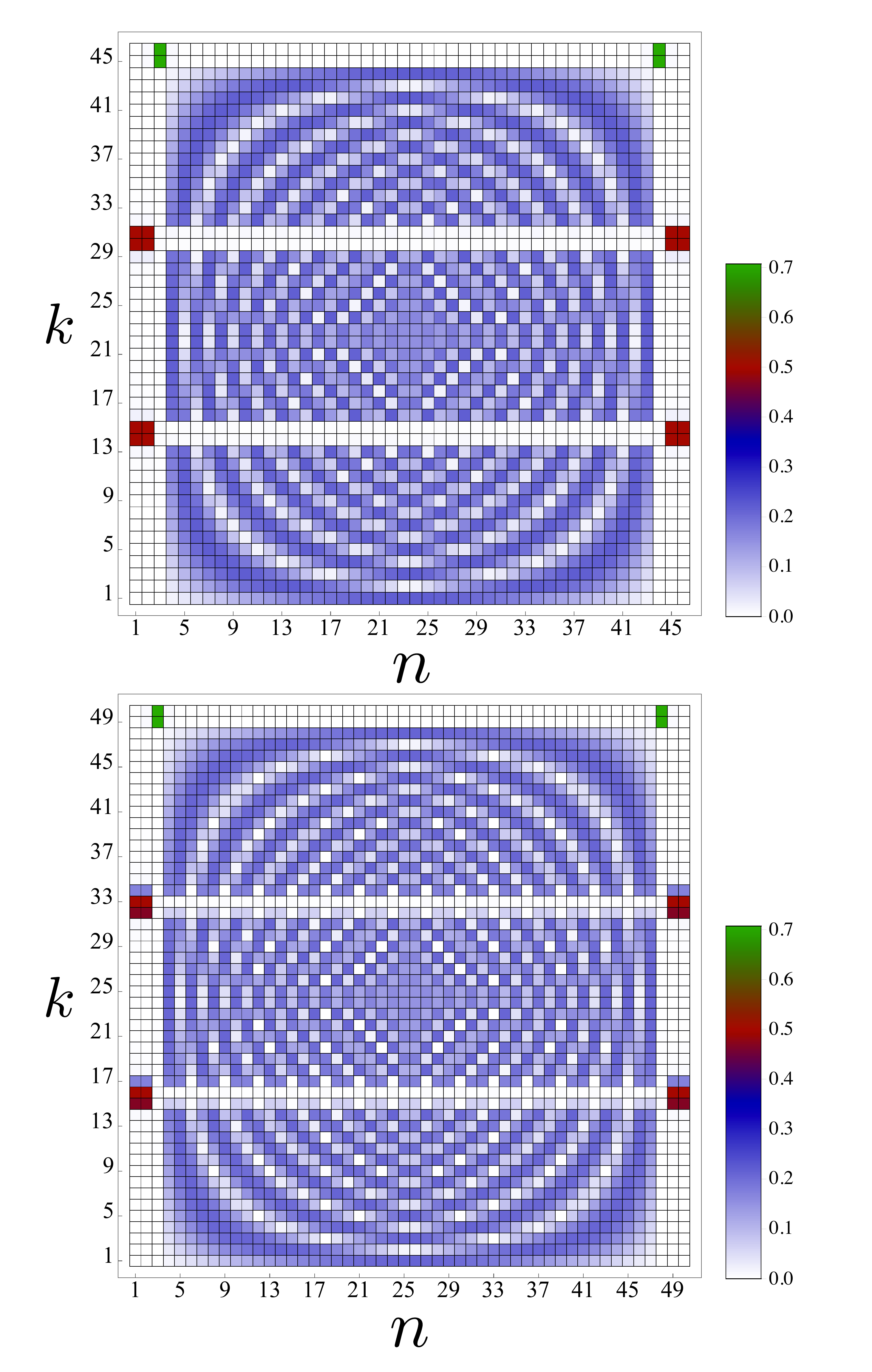}
\caption{(Color online) Density plot of $|a_{kn}|$ versus $k$ and
$n$ for $N=46$ (left) and $N=50$ (right) with $h=100 J$. In the
case of $N\neq 3n-1$ (left panel) it is clearly shown that the
spatial distribution of the eigenvectors exhibits the presence of 4
quadri-localized energy eigenstates; whereas, in the case of
$N=3n-1$ (right panel), two more eigenstates appear with non
negligible amplitudes also on sites $\mathcal{S}$ and
$\mathcal{R}$. Their presence gives rise to a more complicated
dynamical behavior of the fidelity because of the larger number of
degrees of freedom (and frequencies) effectively entering the time
evolution of $\ket{\Psi(0)}$. }
  \label{F.local}
\end{figure}

By writing $H_1=\sum_{k=1}^N \varepsilon_k
\ket{\varepsilon_k}\!\!\bra{\varepsilon_k}$ with the eigenvalues
taken in increasing order, the localized states are labelled by
$k=\{Q[N,3]{-}1,Q[N,3],N{-}Q[N,3]{-}1,N{-}Q[N,3]\}$, where $Q[a,b]$
denotes the quotient. In the following we will refer to these
states by $\{q_i\}$, $i=1,2,3,4$. In Fig.~\ref{F.local} an
instance of such localized structure of the eigenstates is given
for $N=46$ and $h=100 J$. We observe that there are 4 eigenstates,
labelled by $k=14,15,30,31$, that are quadri-localized on the
edges, i.e. at sites $n=1,2,45,46$. Besides these four
eigenstates, another two are bi-localized on the barrier qubits at
sites $n=3,44$; whereas the remaining ones are extended states
with negligible amplitudes on the senders, the receivers, and the
barriers. As a consequence, the contribution to the dynamics of
$\ket{\Psi(0)}$ of these extended states is negligible up to order
$O(h^{-1})$. In the case of $N=3n-1$, on the other hand, two
additional extended eigenstates appear, labelled by
$k=\{Q[N,3]+1,N-Q[N,3]-2\} $ with a non-negligible value of
$a_{kn}$  for $n=1,2,N-1,N$, as shown in Fig.~\ref{F.local}  for
the case of $N=50$. We will refer to the latter as extended
edge-localized states. As a consequence, there are more
eigenstates taking part in the time evolution of the initial state
and, although, high-quality 2-QST is still attainable, the
clear-cut analysis we will give below is, to some extent,
complicated by the their presence. Therefore, in the following, we
will first consider spin chains of length $N\neq3n-1$

\subsection{Rabi-like 2-QST}
Because the quadri-localized states come as a result of the small
effective coupling of $\mathcal{S}$ and $\mathcal{R}$ to the
quantum channel $\Gamma$ (due to the energy mismatch with the
connecting spins at sites $3$ and $N-3$), their energies and
eigenstates can be approximated by 1st-order degenerate
perturbation theory and read
\begin{eqnarray}\label{E.perturbation}
\varepsilon_{q_i}&=&\epsilon^{(0)}_{q_i}+\epsilon^{(1)}_{q_i}, \nonumber \\
\ket{\varepsilon_{q_i}}&=&\!\!\!\!\!\!\!\!\sum_{n=\{1,2,N-1,N\}}\!\!\!\!\!\!\!\!a_{q_i n}\ket{n}~,~|a_{q_i n}|{\simeq}\frac{1}{2},
\end{eqnarray}
where $i=1,2,3,4$.
Notice that the coefficients of the eigenvectors obey the parity
relation $a_{kn}{=}(-1)^{k{+}1}a_{k N{+}1{-}n}$ because of the
mirror symmetry of the model~\cite{BanchiV13}.

Exploiting the quadratic nature of Eq.~(\ref{E.Hfermrealspace}),
we can reduce the two-particles transfer amplitude to
single-particle ones by means of the relation given in
Ref.~\cite{Wangetal12, WangBGS11,nPST2}
\begin{equation}\label{E.2to1}
g_{nm}^{rs}{=}  \begin{vmatrix}

        f_n^r & f_n^{r{+}1} & \cdots & f_n^{s{-}1} & f_n^s \\

            f_{n{+}1}^r &  \ddots  & & & f_{n{+}1}^s \\

      \vdots&   & & &\vdots \\
 f_m^r & \cdots  & & & f_m^s \\

    \end{vmatrix}.
\end{equation}
Moreover, mirror symmetry~\cite{BanchiV13,mirror} implies
$|f_1^{r}|=|f_2^{r{+}1}|$, and perturbation theory allows to
retain only the transition amplitudes between the senders and the
receivers. Working out all these simplifications  the average
fidelity given by Eq.~(\ref{E.Fidelitygen}) reduces to the
approximate expression
\begin{widetext}
\begin{eqnarray}\label{E.Fidapp}
\bar{F_a}(t)&=&\frac{1}{4}+\frac{10}{54}Re\left[f_1^{N-1}\right]+\frac{7}{54}Re\left[\left(f_1^{N-1}\right)^2\right]+\frac{12}{54}|f_1^{N-1}|^2+\frac{2}{54}|f_1^{N}|^2\nonumber \\
&&+\frac{10}{54}|f_1^{N-1}|^2Re\left[f_1^{N-1}\right] -\frac{10}{54}Re\left[f_1^{{N-1}^*}f_1^{N}f_2^{N-1}\right]-\frac{7}{54}Re\left[f_1^{N}f_2^{N-1}\right],
\end{eqnarray}
\end{widetext}
which we remind to be correct up to order $O\left(h^{-1}\right)$.

Eq.~(\ref{E.Fidapp}) will be the starting point of the following
analysis in which we will evaluate both the maximum achievable
fidelity and the optimal transfer time. To start with, let us
notice that $\bar{F_a}(t)$ only depends on the three complex
variables $f_1^{N{-}1},f_1^{N}$, and $f_2^{N{-}1}$, which obey the
constraints
\begin{eqnarray}\label{E.Constr}
0&\leq&|f_1^{N-1}|^2+|f_1^{N}|^2\leq1,\nonumber \\
0&\leq&|f_1^{N-1}|^2+|f_2^{N-1}|^2\leq1,
\end{eqnarray}
because of $\sum_{n=1}^N |f_i^{n}|^2=1$,  for all $i\in
\mathcal{K}$ coming from the conservation of $M^z$. Although
$f_1^{N{-}1}(t),f_1^{N}(t)$, and $f_2^{N{-}1}(t)$ are
complex-valued functions of time,  $\bar{F_a}(t)$ is a real-valued
bounded function, which, taking $\mbox{Re}[f_s^r(t)]$ and
$\mbox{Im}[f_s^r(t)]$ as independent, becomes a function of six
real-valued bounded functions. Therefore, standard Lagrangian
multiplier methods can be applied in order to search for the
absolute maximum of $\bar{F_a}(t)$ within the boundaries given by
Eqs.~(\ref{E.Constr}). It turns out that the maximum of
$\bar{F_a}(t)$ is given by the conditions $Re[f_1^{N-1}(t)]=1$ and
$Re[f_1^{N}(t)]=Re[f_2^{N-1}(t)]=Im[f_s^r(t)]=0$ (the latter
following from the former due to the conservation of $M^z$) and
amounts to $\bar{F_a}(t)=\frac{35}{36}\simeq 0.97$. We found that
$\bar{F_a}(t)$ does not achieve the maximum possible value of $1$
just because it is an approximate expression for the fidelity: in
fact, if the values obtained above for the transition amplitudes
are used in the exact expression of the average fidelity given by
Eq.~(\ref{E.Fidelitygen}), we obtain $\bar{F}(t)=1$.

The next step is to find the time $t^*$ at which the transition
amplitude reaches the optimal values for the 2-QST. To do this, we
can maximize the function $Re[f_1^{N-1}(t)]$ (or equivalently, due
to mirror-symmetry, $Re[f_2^{N}(t)]$) whose time evolution is
generated by the Hamiltonian given in Eq.~(\ref{E.HamBus}). This
will fix the transfer time $t^*$.

Although $\bar{F_a}(t)$ is an highly oscillating function because
of the presence of many frequencies in the transition amplitudes,
it is possible  to find the transfer time $t^*$ of the 2-QST
protocol in a relatively simple way, as we will outline in detail
in the following, for the case of even $N{\neq} 3n{-}1$.

By exploiting Eqs.~(\ref{E.perturbation}) and by means of
elementary trigonometric identities, the term $Re[f_1^{N{-}1}]$
can be expressed as
\begin{eqnarray}\label{E.f1N}
Re[f_1^{N{-}1}]&=&Re[\sum_{k{=}1}^N e^{- i \varepsilon_k t} a_{k1}a_{kN{-}1}] \nonumber \\
                     &\simeq& Re[\sum_{i{=}1}^4 e^{- i \varepsilon_{q_i} t} a_{q_i 1}a_{q_i N{-}1}],
\end{eqnarray}
which, for even $N$, becomes
\begin{eqnarray}\label{E.f1Neven}
Re[f_1^{N-1}]&\simeq&(-1)^{Mod[N,3]{+}1}\times \nonumber \\
&&\times \left(\sin \omega_0^- t\cos \omega_0^+ t   \sin \omega_1^- t  \cos \omega_1^+ t \right. +\nonumber\\
&&\left. +\cos \omega_0^- t \sin \omega_0^+ t  \cos \omega_1^- t  \sin \omega_1^+ t\right),
\end{eqnarray}

where
\begin{eqnarray}\label{E.Freq}
\omega_0^{\pm}&=&\left|\frac{\omega^{\pm}_{14}+\omega^{\pm}_{23}}{2}\right|, \nonumber\\
\omega_1^{\pm}&=&\left|\frac{\omega^{\pm}_{14}-\omega^{\pm}_{23}}{2}\right|, \nonumber\\
\omega_{ij}^{\pm}&=&\frac{\varepsilon_{q_{i}}\pm\varepsilon_{q_{j}}}{2},
\end{eqnarray}
and $Mod[a,b]$ is the modulus function.

\begin{figure}[h!]
\includegraphics[width=\columnwidth]{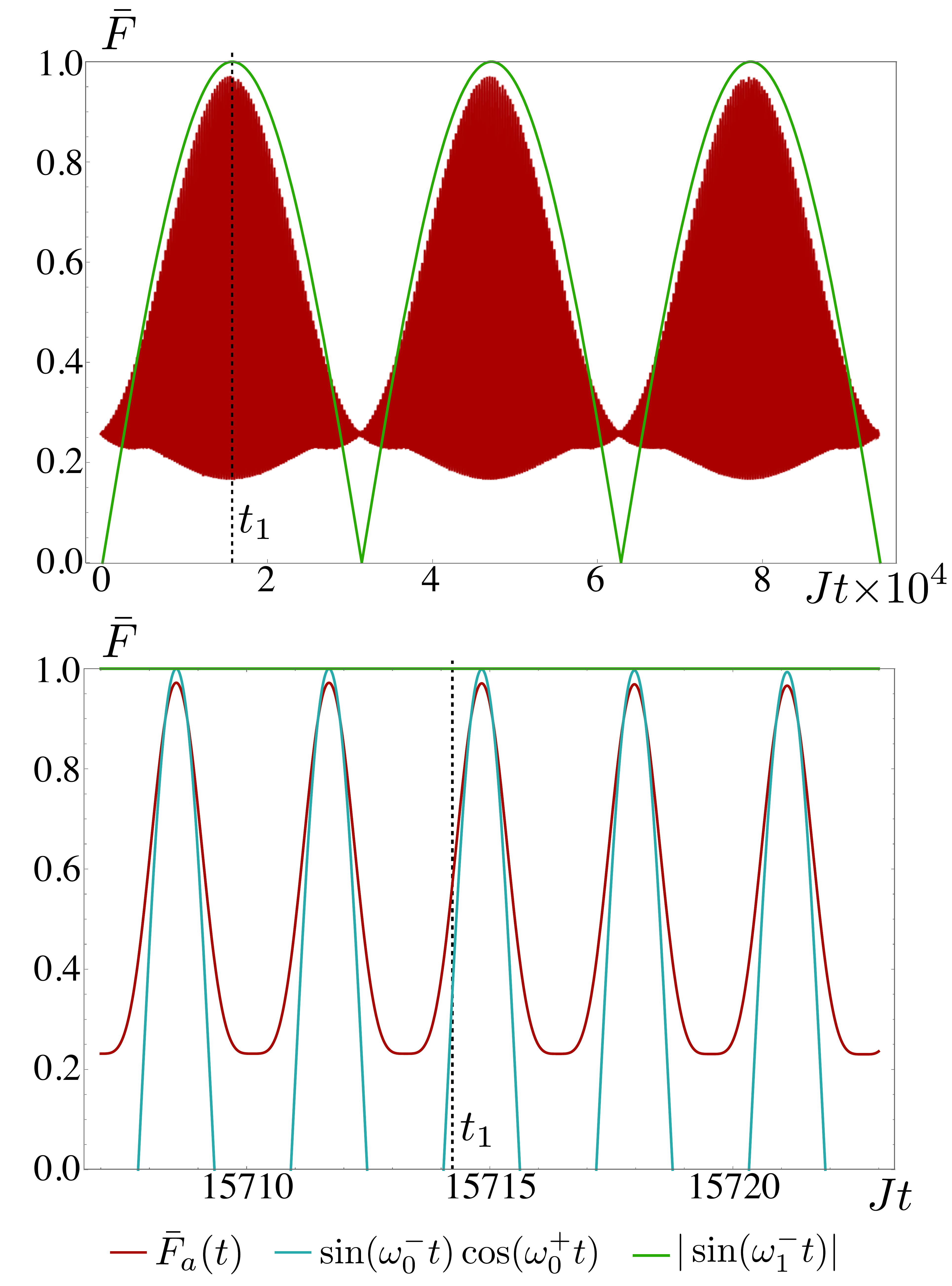}
\caption{(Color online) (left) Approximated average fidelity
$\bar{F}_a(t)$, and, for comparison, $\left| \sin \omega_1^- t
\right |$ shown as a function of time. Notice that $F_a$ is so
rapidly oscillating that it appears to fill the entire red region
in the plot. The sinusoidal function approximately gives an
envelop of the fidelity with the same periodicity, which easily
allows us to identify several \textit{reading} windows for the
2-QST. (right) Full expression for the average fidelity
$\bar{F}_a(t)$ (red), $\sin \omega_0^- t\cos \omega_0^+ t $ (blue)
and $\sin \omega_1^- t$ (green) shown as a function of time around
the optimal transfer time $t_1$, denoted by the vertical dashed
line. Notice that $|\sin \omega_1^- t| \simeq 1$ and that the
fidelity attains very high values near $t_1$ which recur several
times, with a frequency of order of $J^{-1}$, allowing again for
several reading windows. }
  \label{F.timescales}
\end{figure}

Having defined in such a way the frequencies that enter the
dynamics of $Re[f_1^{N{-}1}]$, it turns out that
$\omega_0^-{\gg}\omega_0^+{\gg}\omega_1^-{\gg}\omega_1^+$, which
immediately sets a time scale for the 2-QST. In fact, as
$\omega_1^+ t {\ll} 1$, we can focus only on the first summand of
the RHS in Eq.~\ref{E.f1Neven}, namely $\sin \omega_0^- t \cos
\omega_0^+ t   \sin \omega_1^- t$. On the same footing, the next
time scale is given by $\omega_1^-$, which implies that the
maximum  of $Re[f_1^{N{-}1}](t)$ has to be found in the
neighborhood of $t_1{=}\frac{\pi}{2 \omega_1^-}$, which, hence,
approximately gives the optimal time $t^*$. As a result, the
transfer time $t^*$ is found by solving
\begin{equation}\label{E.time}
t^*=\begin{cases}
\max\left[\sin \omega_0^- t\cos \omega_0^+ t\right]&~~ \mbox{if }\, Mod[N,3] \mbox{ is }\, \textit{odd}\\
\min \left[\sin \omega_0^- t\cos \omega_0^+ t\right]&~~ \mbox{if }\, Mod[N,3] \mbox{ is }\,  \textit{even}
\end{cases}
\end{equation}
and by choosing the  solution closest to the time $t_1$. A
graphical representation of these timescales is given in
Fig.~\ref{F.timescales}, which also allows us to put forward a
simple physical interpretation. In fact, from the left panel of
Fig.~\ref{F.timescales}, $\sin \omega_1^- t$  can be seen as an
approximate envelope for the fidelity of the transfer process,
whereas $\omega_0^-\simeq2 J$  gives the time-scale for the (very
rapid) bouncing of the excitation back and forth between the two
receivers. These oscillations occur because of the direct coupling
between the two receiver spins and could be eliminated if this
coupling is switched off.

\begin{figure}[h!]
\includegraphics[width=\columnwidth]{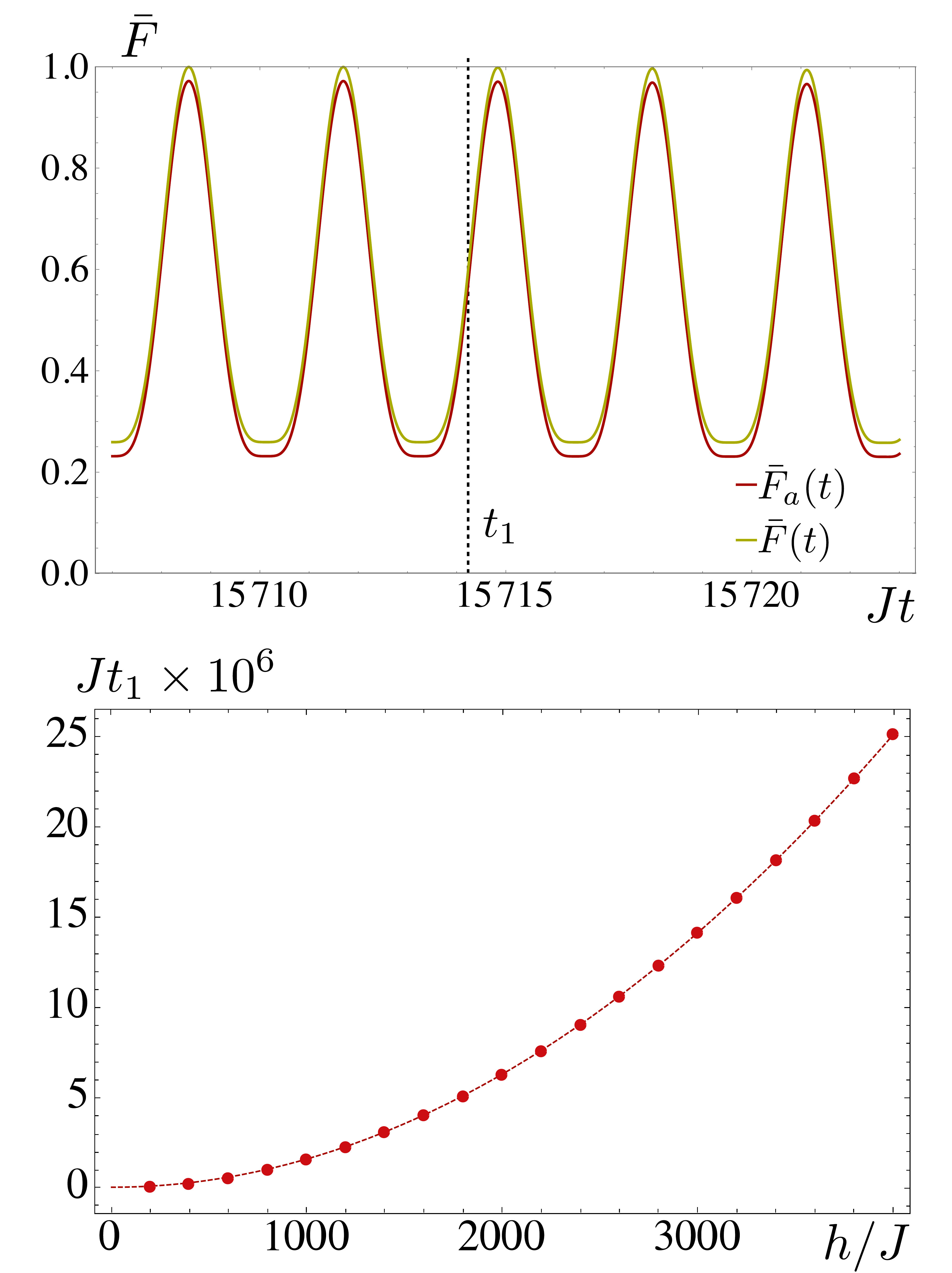}
\caption{(Color online) (left) Exact average fidelity $\bar{F}(t)$
(yellow) and approximate fidelity $\bar{F}_a(t)$  (red) versus $t$
around $t_1$. Notice that $\bar{F}(t)>\bar{F}_a(t)$ is always
fulfilled and that, at the optimal time $t^*$ almost perfect
quantum state transfer is achieved, with $\bar{F}(t)\simeq 0.999$.
(right) Transfer time $t^*$ versus $h$ for $N=30$: the
function $f(h)=\frac{\pi}{2}h^2$ (dashed line) fits perfectly with
the numerical data for $t^*$. The range of values of the magnetic
field $h$ is such that $\bar{F}_a(t)>0.95$. }
  \label{F.comparison}
\end{figure}

This bouncing occurs many times (see the right panel of
Fig.~\ref{F.timescales}) before the excitation slowly leaves these
two sites to go back towards the senders. In
Fig.~\ref{F.comparison} we compare the approximate value of the
fidelity given by Eq.~(\ref{E.Fidapp}) with the exact result of
Eq.~(\ref{E.Fidelitygen}): it is shown that
$\bar{F}(t){>}\bar{F}_a(t)$ attaining values as high as 0.999. We
checked, up to computational accessibility, that this holds true
for every $N$.

The Rabi-like half-oscillation time $t_1$, giving an approximate
value of the optimal time for the excitation transfer from
$\mathcal{S}$ to $\mathcal{R}$, can be obtained by using standard
degenerate time-independent perturbation theory to evaluate the
relevant energy eigenvalues. Here we report the energy corrections
for the dynamically relevant states given in
Eq.~(\ref{E.perturbation}) up to the first order in $h^{-1}$.
Denoting by $x_i$, with $x_1{>}x_2{>}x_3$, the solutions of ${-}x^3{-}hx^2{+}2 x{+}h{=}0$, and defining the parameters $\beta_{1,2}=h{+}x_{1,2}$,
$\alpha=x_{1,2}^2{+}h x_{1,2}{-}1$, and
$\gamma=\left(2\left(\alpha_{1,2}^2{+}\beta_{1,2}^2{+}1\right)\right)^{-\frac{1}{2}}$,
we obtain
\begin{equation}\label{E.perturbationsol}
\varepsilon_{q_1}=\lambda_1^-~,~\varepsilon_{q_2}=\lambda_1^+~,~\varepsilon_{q_3}=\lambda_2^+~,~\varepsilon_{q_4}=\lambda_2^-~,
\end{equation}
where
\begin{equation}\label{E.perturbationsolsol}
\lambda_{1,2}^{\pm}=2\left(z_{1,2}+\frac{\gamma_{1,2}^2}{N-5}\sum_{k=1}^{N-6}\frac{\left(\left(1\pm \cos k \pi\right) \sin\frac{k \pi}{N-5}\right)^2}{z_{1,2}+2 \cos\frac{k \pi}{N-5}}\right).
\end{equation}
Finally, using Eqs.~(\ref{E.Freq}) and~(\ref{E.perturbationsol}),
we obtain the approximate transfer time $t_1{\simeq} /\left
[2\left(\lambda^-_1{-}\lambda^+_1{+}\lambda^-_2{-}\lambda^+_2\right)\right
]$. We also find that $t_1$ scales quadratically with the magnetic
field intensity and that it is independent of $N$,
$t_1{=}\frac{\pi}{2}h^2{+}(-1)^{Mod[N,3]}\frac{Mod[N,2]\pi}{2}h$, as
reported in Fig.~\ref{F.comparison} for the case of $N{=}30$. Notice
that in 1-QST schemes in which the magnetic field is applied
directly on the sender and the receiver, the transfer time scales
exponentially both with the length of the chain $N$ and with the
magnetic field\rq{}s intensity $h$~\cite{Rabi-like3,Rabi-like5}.

A similar procedure for odd $N{\neq} 3n-1$ yields for the transfer
amplitude of Eq.~(\ref{E.f1N})
\begin{eqnarray}\label{E.f1Nodd}
Re[f_1^{N-1}]&\simeq&(-1)^{Mod[N,3]}\times \nonumber \\
&& \times \left(\cos \omega_0^- t\sin \omega_0^+ t   \sin \omega_1^- t  \cos \omega_1^+ t \right.\nonumber\\
&&\left. +\sin \omega_0^- t \cos \omega_0^+ t  \cos \omega_1^- t  \sin \omega_1^+ t\right),
\end{eqnarray}
and the optimal transfer time $t^*$ can be found via a double
step, with a recipe similar to the one discussed above. First we
determine $t_2$, defined as the solution of $\sin \omega_0^+
t{=}(-1)^{Mod[N,3]}$ which is closest to $t_1{=}\frac{\pi}{2
\omega_1^-}$,  and then $t^*$ is given by the solution of $\cos
\omega_0^- t{=}1$ which is closest to $t_2$. Notice that this is not
different from the previous procedure; indeed, there are many ways
to recast Eq.~(\ref{E.f1N}) by combining the energies
$\varepsilon_{q_{i}}$, and the fact that we obtained an apparently
different method for $t^*$ in the Eqs.~(\ref{E.f1Neven}) and
(\ref{E.f1Nodd}), for even and odd $N$, respectively, is due to
the fact that we kept unchanged the definitions of the frequencies
given by Eq.~(\ref{E.Freq}) to avoid a confusing re-labelling.

\subsection{Quasi Rabi-like 2-QST}
Let us now go back to the case of spin chains of length $N{=}3n{-}1$,
that was left out of the previous analysis. In this case, two
additional edge-localized extended states are found, whose
presence hinders the clear Rabi-like oscillations of the
excitation between $\mathcal{S}$ and $\mathcal{R}$, exhibited in
the case of $N{\neq} 3n{-}1$ by the sinusoidal function
$\sin\omega_1^- t$ and reported in the previous subsection and in
Fig.~\ref{F.timescales}. Nevertheless, since the number of
eigenstates (eigenenergies) to be included in the sum given in
Eq.~\ref{E.f1N} increases just by two, an analysis similar to the
one performed above can still be carried out. We dub the
transmission process in this case as {\textit{quasi}} Rabi-like
2-QST.

In fact, the approximated expression for $Re[f_1^{N-1}(t)]$ given
in Eq.~(\ref{E.f1N}) now becomes
\begin{eqnarray}\label{E.f1NN}
Re[f_1^{N{-}1}] &\simeq& Re[\sum_{i{=}1}^6 e^{- i \varepsilon_{q_i} t} a_{q_i 1}a_{q_i N{-}1}] =\nonumber\\
&=&\pm 2 \left(c_1 \cos\omega_{14}^+t \cos\omega_{14}^-t+ \right.  \nonumber \\
&&-c_2\cos\omega_{25}^+t \cos\omega_{25}^-t+\nonumber \\
&& +\left. c_3\cos\omega_{36}^+t \cos\omega_{36}^-t   \right),
\end{eqnarray}
with the $\pm$ sign holding for even (odd) $N$, respectively, and
where we exploited the mirror-symmetry and used $1$st-order
degenerate perturbation theory relations
\begin{eqnarray}
c_1&=& a_{11} a_{1N-1}\simeq a_{41} a_{4N-1}\simeq\frac{1}{4}-\frac{3}{2 N-1},\\
c_2&=& a_{21} a_{2N-1}\simeq a_{51} a_{5N-1}\simeq\frac{1}{4},\\
c_3&=& a_{31} a_{3N-1}\simeq a_{61} a_{6N-1}\simeq \frac{3}{2N-1}.
\end{eqnarray}

Using the approximations $\omega_{14}^-\simeq \omega_{25}^- \simeq
\omega_{36}^-\simeq -2$ and $\omega_{36}^+\simeq 0$, yields
\begin{eqnarray}
Re[f_1^{N{-}1}] \simeq \pm \cos 2 t \left(4 c_3 \sin\omega_{14}^+  t-\sin^2 \omega_{14}^+ t\right).
\end{eqnarray}
Once again, it is possible to identify two different processes:
the slow quasi-Rabi-like oscillations of the excitation between
$\mathcal{S}$ and $\mathcal{R}$, having a time scale ruled by
$\omega_{14}^+$, and the fast oscillations of the excitation
within $\mathcal{R}$ ($\mathcal{S}$) triggered by $J$ and
described by the term $\cos 2 t$. Although the additional states
complicate somehow the expression of $Re[f_1^{N{-}1}]$, they also
provide a clear advantage as far as the transfer time is
concerned. Indeed the time $t_1$ is now linear in the magnetic
field and hence 2-QST occurs faster w.r.t. chains of length $N{\neq}3n{-}1$. 
Let us remind that a similar phenomenon takes place in
1-QST protocols where the sender and the receiver are weakly
coupled to the chain either because of smaller bond
strengths~\cite{Rabi-like4} or because of a strong magnetic field
acting on barrier qubits~\cite{Rabi-like6}. Indeed, in those cases
the QST time for even- and odd-length chains too scales,
respectively, quadratically and linearly with the
perturbation\rq{}s intensity.

\begin{figure}
\includegraphics[width=\columnwidth]{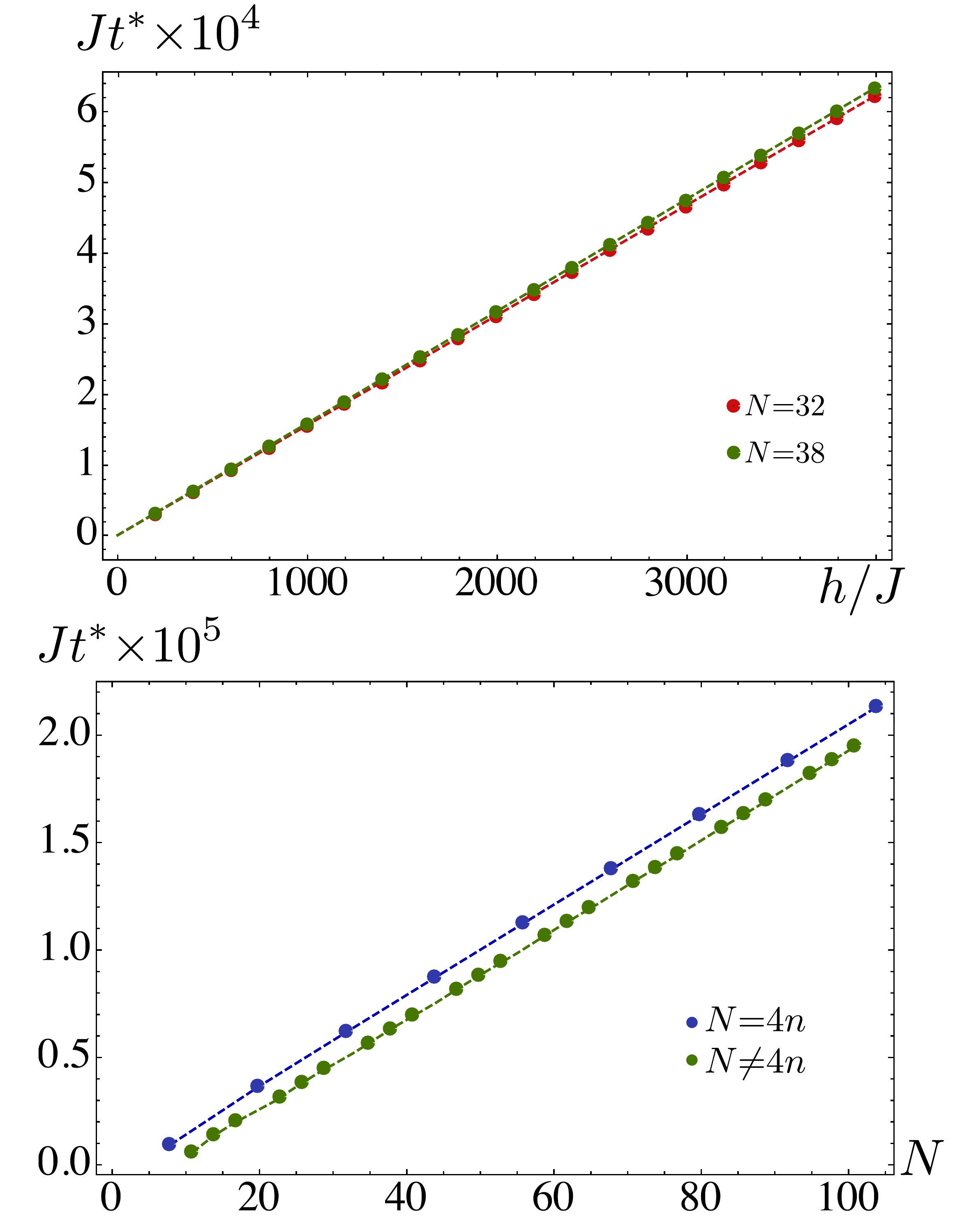}
\caption{(Color online) (left) Transfer time $t^*$ for the case of
spin chain\rq{}s of length $N{=}3n{-}1$ as a function of $h$. The time
$t^*$ increases linearly with $h$, although with slightly
different coefficients depending on wether $N$ is divisible by $4$
or not. In the figure the two cases are exemplified by $N{=}32$ and
$N{=}38$, respectively. (right) Transfer time $t^*$ for the case of
spin chain\rq{}s of length $N{=}3n{-}1$ with magnetic field $h{=}4000$
as a function of $N$. The time increases linearly with $N$, with
clearly distinct coefficients depending on the divisibility by 4
of $N$.}
  \label{F.tempi_lin}
\end{figure}

\begin{figure}\centering
\includegraphics[width=1\columnwidth]{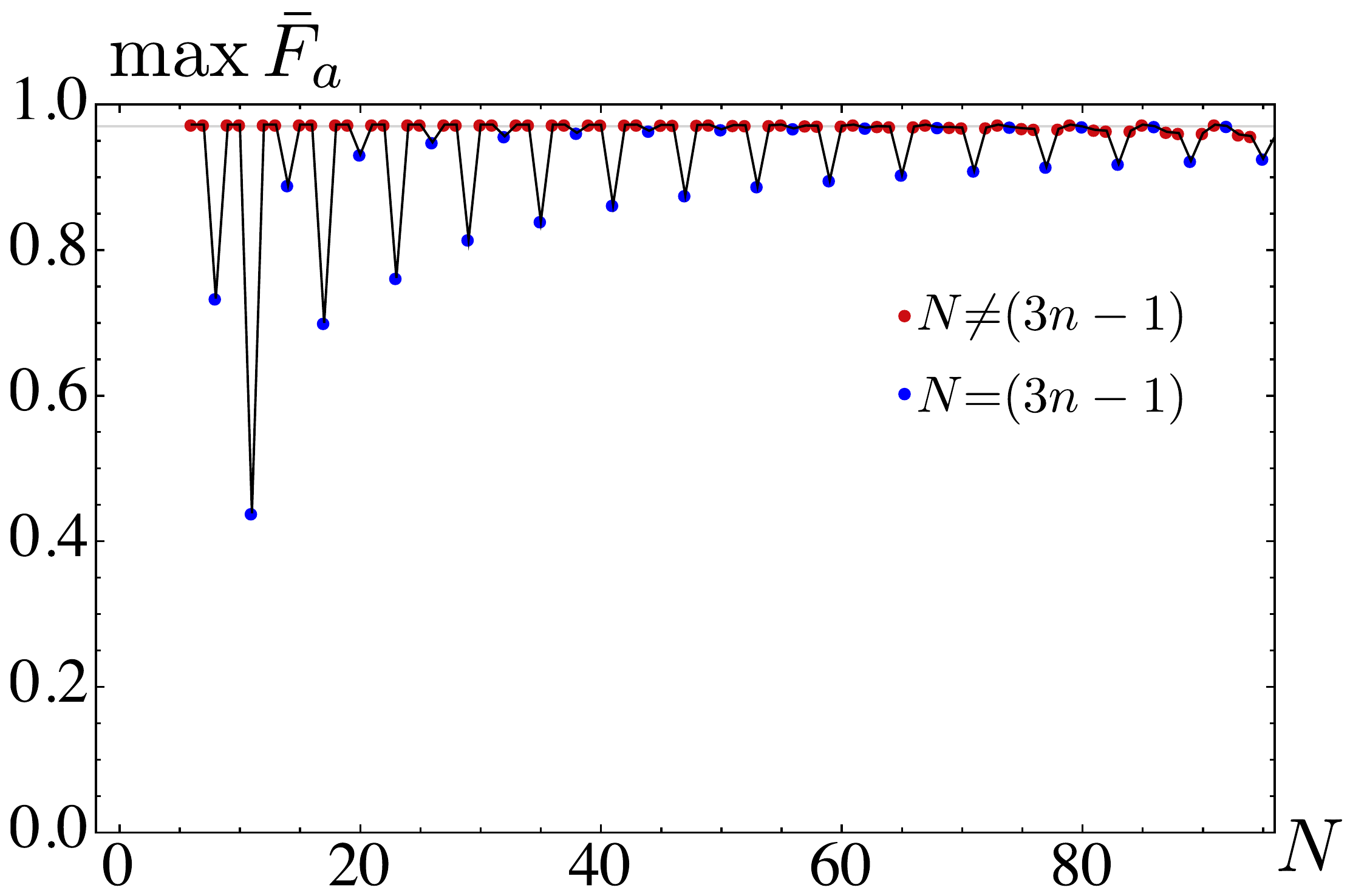}
\caption{(Color online) Maximum of the approximate fidelity
$\bar{F}_a(t^*)$ obtained by choosing $h{=}4000$. The red points are
the Rabi-like 2-QST and the maximum does not depend on $N$, the
blue point are for $N{=}3n{-}1$ and even $N$ performs better than odd
$N$. For $N{\gg}1$, both of the curves converge to the $N{\neq}3n{-}1$
case.}
  \label{F.fmax}
\end{figure}

In Fig.~\ref{F.fmax}, we summarize the main result of the two
previous subsections, namely the possibility to transfer with
high-fidelity an arbitrary quantum state of two qubits by means of
a linear spin-$\frac{1}{2}$ chain with strong magnetic fields on
the barrier qubits. It is shown that for $N{\neq}3n{-}1$, a fidelity
close to unity can be achieved regardless of $N$ (provided strong
enough magnetic fields are applied at the barrier sites) although
in a time that increases quadratically with $h$. For $N{=}3n{-}1$,
instead, the quality of the transfer depends on whether $N$ is
divisible by 4 (higher $\bar{F}_a(t^*)$) or not (lower
$\bar{F}_a(t^*)$), but the transfer is achieved in a time that
scales only linearly with both $h$ and $N$. Nevertheless, the
differences amongst all these cases fade away for $N{\gg} 1$, where
all curves collapse.

Since the average fidelity is not \textit{identically} one, one
could imagine that there exist specific input states that are
transferred with a relatively poor quality. This is not the case,
and in order to dispel such a doubt, we evaluated also the worst
case fidelity~\cite{ballistic1} and found that the minimum
state-dependent fidelity, evaluated by means of
Eq.~\ref{E.Fidelity}, remains close to the average one up to the
fourth digit, i.e. $F_{min}\simeq 0.999$.

To conclude this Section we stress that the probability to find
the excitations inside the quantum channel $\Gamma$, evaluated by
$\sum_{n=3}^{N-2}\left(\left|f_1^n(t)\right|^2+\left|f_2^n(t)\right|^2\right)$,
is radically different depending on whether $N{\neq} 3n{-}1$ or not.
In the former case it is of the order of $O(h^{-1})$ because
$\Gamma$ acts as a mere physical connector~\cite{Rabi-like3,
Rabi-like4, Rabi-like9} entering the dynamics only virtually.
On the contrary, when $N{=}3n{-}1$, two extended states with a
non-negligible overlap on the senders and the receivers come into
play, meaning that the excitations can be actually found inside
the channel $\Gamma$. As a consequence, the effect of a
dissipative coupling of an environment eventually acting only on
$\Gamma$ has a negligible influence only for the Rabi-like QST
whereas for the quasi-Rabi-like one the quality could be severely
degraded, especially for short chains. For long chains, on the
other hand, since the overlap with the extended state localized
also on the edges scales as $c_3\sim O(N^{-1})$, the degrading
effect becomes negligible. On the other hand, the presence of
disorder in the couplings or in a magnetic field acting on the
quantum channel $\Gamma$ should have a negligible influence  too
on the efficiency of the 2-QST protocol we are proposing here,
regardless of $N$, as long as the spatial distribution of the
eigenvectors depicted in Fig.~\ref{F.local} is not significantly
affected~\cite{disorder1,disorder2,disorder3}.

\section{Conclusions}\label{S.Concl}

In this paper, we derived an expression for the average fidelity of the
quantum state transfer of two qubits through a spin-$\frac{1}{2}$ chain, providing that 
the $z$-total angular momentum is conserved 
and all the spins are initially aligned.  This general expression, relating the average
fidelity explicitly to one- and two-particle transition
amplitudes, may result useful in investigating the two qubit QST
properties of a wide range of physical models displaying the
above-mentioned characteristics.

Furthermore, we discussed a specific case, obtained by extending a
Rabi-like protocol, widely adopted for QST of single qubits, to
the non-trivial case of the QST of two qubits, where the senders
and receivers pairs are located at each end of a one dimensional
spin-$\frac{1}{2}$ chain with $XX$-Heisenberg type
nearest-neighbor interactions. The presence of strong magnetic
fields on the two sites closest to the sender and receiver
(barrier qubits), allowed us to obtain a faithfully transfer for
an arbitrary two-qubits quantum state. We characterized the
quality of the 2-QST by using first-order degenerate perturbation
theory; thus providing, apart from a clear-cut physical
interpretation of the multi-excitation dynamics yielding
high-quality 2-QST, also an approximate analytical expression for
the time transfer. The latter is found to increase  linearly or
quadratically with the magnetic field intensity $h$ depending on
the spin chain length. Moreover, we have verified that the
worst-case fidelity of the QST remains almost unchanged with
respect to the average one, i.e., $\simeq 0.999$.

Two final comments are in order. Since it is straightforward that
the 4-dimensional Hilbert space of the senders can be also
employed to encode an $n<4$ qudit to be transferred to the
receivers, the scheme we propose can be adapted to the transfer of
qutrits (qubits) encoded in arbitrary three (two) orthogonal
quantum states of the senders. Finally, the question if the
barrier scheme discussed here is useful in order to perform QST of
an arbitrary number of qubits will be left to future
investigations.

\section{Acknowledgement}
{TJGA and GMP acknowledge the EU Collaborative Project TherMiQ (Grant Agreement 618074). 
SL and GMP acknowledge support by MIUR under PRIN 2010/11
TJGA thanks the International Institute of Physics - UFRN (Natal, Brazil) for the kind hospitality provided during part of this work.
SP is supported by a Rita Levi-Montalcini fellowship of MIUR. SP and TJGA acknowledge partial support from MCTI and UFRN/MEC (Brazil).}

\end{document}